\documentclass[letter,twocolumn]{jpsj3}
\usepackage{amsmath,txfonts,bm}
\newcommand\dc[1]{\dot {\cal #1}}

\newcommand\hb[1]{\hat {\bm #1}}

\DeclareMathOperator\diag{diag}
\title{Theory of Charge and Heat Polarizations with the Keldysh Formalism}
\author{Atsuo Shitade}
\date{\today}
\inst{Department of Physics, Kyoto University, Kyoto 606-8502, Japan}
\abst{%
  We investigate the heat polarization, a heat analog of the charge polarization, by using the gauge-covariant Keldysh formalism.
  In contrast to the charge-heat analogy naively expected, we find that the heat polarization does not appear spontaneously,
  since it consists not only of the heat-transfer contribution but of the heat-generation contribution, leading to the Mott rule.
  Nonetheless, it can be induced by a torsional magnetic field in $(3+1)$-D topological insulators and superconductors,
  which is described by the temporal part of the Nieh-Yan action.
}
\begin{document}
\maketitle
%--- Introduction
The charge polarization (CP) is one of the most important quantities in the condensed matter physics.
In the classical electromagnetism in matters, it is defined by the volume integral of the product of the charge density and the position.
However, the position operator is ill-defined in crystals, and the CP cannot be defined naively.
Furthermore, the CP is not thermodynamically defined by the free-energy perturbation with respect to an electric field.
To overcome these problems, an ingenious formalism was proposed to calculate the temporal integral of the charge current
induced by an adiabatic change in the Hamiltonian~\cite{PhysRevB.47.1651,PhysRevB.48.4442,RevModPhys.66.899}.
The CP is associated with the Berry connection, and is interpreted as the expectation value of the position operator in the Wannier basis.
Many-body extensions were proposed by averaging over boundary conditions~\cite{PhysRevB.49.14202},
introducing the periodic position operator in the periodic boundary condition~\cite{PhysRevLett.80.1800},
and the Green-function formalism~\cite{PhysRevB.84.205137,PhysRevB.88.155121}.

The heat polarization (HP) is a heat analog of the CP, and is a textbook concept to introduce the heat current~\cite{9780306463389}.
However, it has not been formulated in crystals since there are the same fundamental difficulties as in the CP and the heat current was not properly defined.
If the analogy between charge and heat is expected,
it is a natural question whether the HP is possible in inversion-broken insulators or not.

There is only one previous work related to the HP in the context of ($3+1$)-D topological insulators and superconductors (TIs/TSCs),
in which it can be induced by an angular velocity of rotation~\cite{PhysRevLett.108.026802}.
It is an important issue to microscopically derive the effective actions for the topological heat responses in TIs/TSCs.
Unfortunately, it is still difficult because the cross correlation predicted in Ref.~\citen{PhysRevLett.108.026802} is the second-order perturbation effect with respect to gravity,
but we believe that formulation of the HP is the important first step to approach this issue.

In this Letter, we show an alternative derivation of the CP with the Keldysh formalism~\cite{9780521874991,9780521760829,PTP.116.61,1310.8043}.
Now that the gauge structure of heat responses was established based on a theory of gravity~\cite{PhysRev.135.A1505,1310.8043},
we first derive the Berry-phase formula of the HP.
The HP consists of the heat-transfer contribution with the free energy being conserved and the heat-generation contribution.
If we identify the adiabatic parameter, time, to the momentum,
the former corresponds to the Kubo-formula contribution to the thermoelectric Hall conductivity, the latter corresponds to the orbital magnetization,
and hence the total HP vanishes near zero temperature owing to the Mott rule~\cite{0022-3719-10-12-021,PhysRevB.55.2344,PhysRevLett.107.236601}.
Finally we mention that the HP can be induced by a torsional magnetic field in $(3+1)$-D TIs/TSCs,
which is related to their axion electrodynamics through the Wiedemann-Franz law
and is described by the temporal part of the Nieh-Yan action~\cite{nieh:373}.

Hereafter we assign the Latin ($a, b, \dots = {\hat 0}, {\hat 1}, \dots, {\hat d}$) and Greek ($\mu, \nu, \dots = 0, 1, \dots, d$) alphabets
to locally flat and global coordinates, respectively.
We follow the Einstein convention, which implies summation over the spacetime dimension $D = d + 1$ when an index appears twice in a single term.
The Minkowski metric is taken as $\eta_{ab} = \diag (-1, +1, \dots, +1)$.
The Planck constant and the charge are denoted by $\hbar$ and $q$, while the speed of light and the Boltzmann constant are put to $c = k_{\rm B} = 1$.
The upper or lower signs in equations correspond to boson or fermion.

%--- Formalism
In order to calculate the CP/HP,
we consider an adiabatic change in the Hamiltonian, ${\cal H}(X^0)$~\cite{PhysRevB.47.1651,PhysRevB.48.4442,RevModPhys.66.899}.
The dynamics of the Green function in a curved spacetime is determined by the Dyson equation~\cite{1310.8043},
\begin{equation}
  ({\cal L}- {\hb \Sigma}) \star {\hb G}(X, \pi, S)
  = 1, \label{eq:dyson}
\end{equation}
where ${\cal L}$ is the Lagrangian density, and ${\hb G}$ and ${\hb \Sigma}$ are the Green function and the self-energy in the matrix representation,
\begin{align}
  {\hat G}
  = & \begin{bmatrix}
        G^{\rm R} & 2 G^< \\
        0 & G^{\rm A}
      \end{bmatrix}, &
  {\hat \Sigma}
  = & \begin{bmatrix}
        \Sigma^{\rm R} & 2 \Sigma^< \\
        0 & \Sigma^{\rm A}
      \end{bmatrix}. \label{eq:keldysh}      
\end{align}
Among the retarded, advanced, and lesser components indicated by ${\rm R}$, ${\rm A}$, and $<$,
the lesser component is a quantum extension of the distribution function and is important to calculate the thermal expectation values.
$X$ is the center-of-mass coordinate,
$\pi$ is the mechanical momentum obtained by the Fourier transformation of the covariant derivative with respect to the relative coordinate,
and $S$ is the spin.
In the covariant derivative $D_a \equiv h_a^{\phantom{a} \mu} (\partial_{\mu} - i q A_{\mu}/\hbar - i \omega^{ab}_{\phantom{ab} \mu} S_{ab}/2 \hbar)$,
gauge potentials are taken into account such as a vector potential $A_{\mu}$ corresponding to U$(1)$ transformations,
a vielbein $h^a_{\phantom{a} \mu}$ corresponding to spacetime translations,
and a spin connection $\omega^{ab}_{\phantom{ab} \mu}$ to local Lorentz transformations~\cite{9780521278584,9780521877879}.
The star product $\star$ can be constructed in principle from the ordinary product and the Poisson bracket,
\begin{align}
  {\cal P}
  = & (\partial_{X^a} \otimes \partial_{\pi_a} - \partial_{\pi_a} \otimes \partial_{X^a}) \notag \\
  & + (\pi_a \eta_{be} + \eta_{ad} S_{bc} \omega^{cd}_{\phantom{cd} e})
  (\partial_{\pi_e} \otimes \partial_{S_{ab}} - \partial_{S_{ab}} \otimes \partial_{\pi_e}) \notag \\
  & + \eta_{ad} S_{bc} \partial_{S_{ab}} \otimes \partial_{S_{cd}} \notag \\
  & + (q F_{cd} + T^a_{\phantom{a} cd} \pi_a + R^{ab}_{\phantom{ab} cd} S_{ab}/2) \partial_{\pi_c} \otimes \partial_{\pi_d}, \label{eq:poisson}
\end{align}
in which $F_{cd}$, $T^a_{\phantom{a} cd}$, and $R^{ab}_{\phantom{ab} cd}$ are gauge fields called electromagnetic fields, torsion, and Riemann tensors, respectively.
Now that we focus on an adiabatic change in the Hamiltonian,
we expand the star product, the Green function, and the self-energy with respect to the spacetime gradient up to the first order,
\begin{subequations} \begin{align}
  \star
  = & 1
  + \frac{i \hbar}{2} h_a^{\phantom{a} \mu} (\partial_{X^{\mu}} \otimes \partial_{\pi_a} - \partial_{\pi_a} \otimes \partial_{X^{\mu}}) + \dots, \label{eq:star} \\
  {\hb G}
  = & {\hat G}_0 + \frac{\hbar}{2} {\hat G}_1 + \dots, \label{eq:g} \\
  {\hb \Sigma}
  = & {\hat \Sigma}_0 + \frac{\hbar}{2} {\hat \Sigma}_1 + \dots \label{eq:s}
\end{align} \label{eq:expand} \end{subequations}
The subscripts $0, 1$ in the Green function and the self-energy indicate the zeroth and first-order spacetime gradients, respectively,
and the capital letter indicates that the effects of disorder or interactions are taken into account.
We put a vielbein $h^a_{\phantom{a} \mu} = \delta^a_{\phantom{a} \mu}$ when we calculate the charge and heat currents below, but leave it arbitrary for convenience.
By substituting these into the Dyson equation Eq.~\eqref{eq:dyson}, we obtain ${\hat G}_0 = ({\cal L} - {\hat \Sigma}_0)^{-1}$ and
\begin{align}
  {\hat G}_1
  = & {\hat G}_0 {\hat \Sigma}_1 {\hat G}_0 + i h_a^{\phantom{a} \mu} \notag \\
  & \times ({\hat G}_0 \partial_{X^{\mu}} {\hat G}_0^{-1} {\hat G}_0 \partial_{\pi_a} {\hat G}_0^{-1} {\hat G}_0 - (X^{\mu} \leftrightarrow \pi_a)). \label{eq:g1}
\end{align}
In order to calculate the thermal expectation values, it is important to extract the lesser Green function.
We use the equilibrium condition, $G_0^< = \pm (G_0^{\rm R} - G_0^{\rm A}) f(-\pi_{\hat 0})$,
and decompose the lesser Green function and the self-energy as
\begin{subequations} \begin{align}
  G_1^<
  = & G_1^{< (0)} f + G_1^{< (1)} f^{\prime}, \label{eq:g1l} \\
  \Sigma_1^<
  = & \Sigma_1^{< (0)} f + \Sigma_1^{< (1)} f^{\prime}. \label{eq:s1l}
\end{align} \label{eq:1l} \end{subequations}
As a result, we obtain the first-order real-time Green functions with respect to the spacetime gradient,
\begin{subequations} \begin{align}
  G_1^{\rm R}
  = & G_0^{\rm R} \Sigma^{\rm R}_1 G_0^{\rm R} + i h_a^{\phantom{a} \mu} \notag \\
  & \times (G_0^{\rm R} \partial_{X^{\mu}} G_0^{{\rm R} -1} G_0^{\rm R} \partial_{\pi_a} G_0^{{\rm R} -1} G_0^{\rm R}
  - (X^{\mu} \leftrightarrow \pi_a)), \label{eq:g1r} \\
  G_1^{< (0)}
  = & \pm (G_1^{\rm R} - G_1^{\rm A}), \label{eq:g10} \\
  \Sigma_1^{< (0)}
  = & \pm (\Sigma_1^{\rm R} - \Sigma_1^{\rm A}), \label{eq:s10} \\
  G_1^{< (1)}
  = & G_0^{\rm R} \Sigma_1^{< (1)} G_0^{\rm A}
  \pm i h_{\hat 0}^{\phantom{\hat 0} \mu}
  [G_0^{\rm R} \partial_{X^{\mu}} G_0^{{\rm R} -1} (G_0^{\rm R} - G_0^{\rm A}) \notag \\
  & - (G_0^{\rm R} - G_0^{\rm A}) \partial_{X^{\mu}} G_0^{{\rm A} -1} G_0^{\rm A}]. \label{eq:g11}
\end{align} \label{eq:perturbx} \end{subequations}
The self-energy $\Sigma_1^{< (1)}$ is determined self-consistently.

%--- Charge polarization
As a demonstration, first we derive the CP by calculating the charge current under an adiabatic change in the Hamiltonian.
The thermal expectation value of the charge current in the Wigner representation is given by
\begin{equation}
  J^{\hat \imath}(X)
  = \pm \frac{i \hbar q}{2} \int \frac{d^D \pi}{(2 \pi \hbar)^D} \tr [v^{\hat \imath} \star {\hb G} + {\hb G} \star v^{\hat \imath}]^<, \label{eq:j}
\end{equation}
where $v^{\hat \imath}$ is the renormalized velocity in disordered or interacting systems.
The change in the CP is given by the temporal integral of the charge current,
\begin{align}
  \Delta P^{\hat \imath}
  \equiv & \int_0^T d X^0 J^{\hat \imath}(X) \notag \\
  = & \pm \frac{i \hbar^2 q}{2} \int_0^T d X^0 \int \frac{d^D \pi}{(2 \pi \hbar)^D} \notag \\
  & \times \tr v^{\hat \imath} [G_1^{< (0)} f(-\pi_{\hat 0}) + G_1^{< (1)} f^{\prime}(-\pi_{\hat 0})]. \label{eq:dp}
\end{align}
Note that the star product is reduced to the ordinary product owing to symmetrization.

Below let us concentrate on the clean and non-interacting limit, ${\hb \Sigma} = 0$.
In this limit, we have
\begin{subequations} \begin{align}
  \Delta P^{\hat \imath}
  = & -\frac{\hbar^2 q}{2} \int_0^T d X^0 \int \frac{d^D \pi}{(2 \pi \hbar)^D} f(-\pi_{\hat 0}) \notag \\
  & \times \tr [g_0^{\rm R} v^{\hat \imath} g_0^{\rm R} {\dc H} g_0^{\rm R} - ({\rm R} \to {\rm A})] - (v^{\hat \imath} \leftrightarrow {\dc H}) \label{eq:dp1a} \\
  & + \frac{\hbar^2 q}{2} \int_0^T d X^0 \int \frac{d^D \pi}{(2 \pi \hbar)^D} f^{\prime}(-\pi_{\hat 0}) \notag \\
  & \times \tr v^{\hat \imath} [g_0^{\rm R} {\dc H} (g_0^{\rm R} - g_0^{\rm A}) - (g_0^{\rm R} - g_0^{\rm A}) {\dc H} g_0^{\rm A}]. \label{eq:dp1b}
\end{align} \label{eq:dp1} \end{subequations}
where $g_0^{\rm R/A} = (-\pi_{\hat 0} - {\cal H} + \mu \pm i \eta)^{-1}$ is the retarded/advanced Green function,
$v^{\hat \imath} = -\partial_{\pi_{\hat \imath}} g_0^{{\rm R/A} -1}$ is the velocity, and ${\dc H} \equiv -\partial_{X^0} g_0^{{\rm R/A} -1}$.
In an adiabatic change, we can define the eigenstates satisfying ${\cal H} | u_{n {\vec \pi} X^0} \rangle = \epsilon_{n {\vec \pi} X^0} | u_{n {\vec \pi} X^0} \rangle$.
By expanding the trace with respect to these states and employing the integral over $-\pi_{\hat 0}$ with the residue theorem,
we obtain two terms from Eq.~\eqref{eq:dp1a},
\begin{subequations} \begin{align}
  \Delta P^{(0) {\hat \imath}}
  = & -q \int_0^T d X^0 \int \frac{d^d \pi}{(2 \pi \hbar)^d} \sum_n \Omega_{n {\vec \pi} X^0}^{\hat \imath} f_{n {\vec \pi} X^0}, \label{eq:dp2a} \\
  \Delta P^{(1b) {\hat \imath}}
  = & \frac{q}{2} \int_0^T d X^0 \int \frac{d^d \pi}{(2 \pi \hbar)^d} \sum_n
  m_{n {\vec \pi} X^0}^{\hat \imath} f^{\prime}_{n {\vec \pi} X^0}, \label{eq:dp2b}
\end{align} \label{eq:dp2} \end{subequations}
and one term from Eq.~\eqref{eq:dp1b},
\begin{subequations} \begin{align}
  \Delta P^{(1a) {\hat \imath}}
  = & -\frac{q}{2} \int_0^T d X^0 \int \frac{d^d \pi}{(2 \pi \hbar)^d} \sum_n
  m_{n {\vec \pi} X^0}^{\hat \imath} f^{\prime}_{n {\vec \pi} X^0}, \label{eq:dp2c} \\
  \Omega_{n {\vec \pi} X^0}^{\hat \imath}
  \equiv & i \hbar \sum_m
  \frac{\langle u_{n {\vec \pi} X^0} | v^{\hat \imath} | u_{m {\vec \pi} X^0} \rangle \langle u_{m {\vec \pi} X^0} | {\dc H} | u_{n {\vec \pi} X^0} \rangle}
  {(\epsilon_{n {\vec \pi} X^0} - \epsilon_{m {\vec \pi} X^0})^2} \notag \\
  & - (v^{\hat \imath} \leftrightarrow {\dc H}), \label{eq:berrycurv} \\
  m_{n {\vec \pi} X^0}^{\hat \imath}
  \equiv & i \hbar \sum_m
  \frac{\langle u_{n {\vec \pi} X^0} | v^{\hat \imath} | u_{m {\vec \pi} X^0} \rangle \langle u_{m {\vec \pi} X^0} | {\dc H} | u_{n {\vec \pi} X^0} \rangle}
  {\epsilon_{n {\vec \pi} X^0} - \epsilon_{m {\vec \pi} X^0}} \notag \\
  & - (v^{\hat \imath} \leftrightarrow {\dc H}), \label{eq:mag}
\end{align} \label{eq:dp3} \end{subequations}
where $\Omega_{n {\vec \pi} X^0}^{\hat \imath}$ and $m_{n {\vec \pi} X^0}^{\hat \imath}$
are the Berry curvature and the ``magnetic moment'' in the $(\pi_{\hat \imath}, X^0)$ space.
In the clean and non-interacting limit, Eqs.~\eqref{eq:dp2b} and \eqref{eq:dp2c} are canceled exactly.
In insulators at zero temperature, we can employ the temporal integral in Eq.~\eqref{eq:dp2a} safely,
and obtain the CP itself~\cite{PhysRevB.47.1651,PhysRevB.48.4442,RevModPhys.66.899} as
\begin{equation}
  P^{\hat \imath}(X^0)
  = q \int \frac{d^d \pi}{(2 \pi \hbar)^d} \sum_n^{\rm occ} A_{n {\vec \pi} X^0}^{\hat \imath}, \label{eq:cp}
\end{equation}
where $A_{n {\vec \pi} X^0}^{\hat \imath}$ is the Berry connection defined by
\begin{equation}
  A_{n {\vec \pi} X^0}^{\hat \imath}
  \equiv i \hbar \langle u_{n {\vec \pi} X^0} | \partial_{\pi_{\hat \imath}} | u_{n {\vec \pi} X^0} \rangle. \label{eq:berrycon}
\end{equation}

%--- Heat polarization
Next we calculate the heat current under an adiabatic change in the Hamiltonian to define the HP.
The thermal expectation value of the conserved heat current is given by~\cite{1310.8043}
\begin{align}
  J_{\rm Qc}^{\hat \imath}(X)
  = & \pm \frac{i \hbar}{2} \int \frac{d^D \pi}{(2 \pi \hbar)^D} \notag \\
  & \times\tr [(-\pi_{\hat 0}) \star {\hb G} \star v^{\hat \imath} + v^{\hat \imath} \star {\hb G} \star (-\pi_{\hat 0})]^<. \label{eq:jqc}
\end{align}
One part of the change in the HP, called the heat-transfer contribution, is given by the temporal integral of the heat current,
\begin{align}
  \Delta P_{\rm Qc}^{\hat \imath}
  \equiv & \int_0^T d X^0 J_{\rm Qc}^{\hat \imath}(X) \notag \\
  = & \pm \frac{i \hbar^2}{2} \int_0^T d X^0 \int \frac{d^D \pi}{(2 \pi \hbar)^D} (-\pi_{\hat 0}) \notag \\
  & \times \tr v^{\hat \imath} [G_1^{< (0)} f(-\pi_{\hat 0}) + G_1^{< (1)} f^{\prime}(-\pi_{\hat 0})]. \label{eq:dpqc}
\end{align}
By the same calculations as in the CP, we find three terms in the clean and non-interacting limit,
\begin{subequations} \begin{align}
  \Delta P_{\rm Qc}^{(0) {\hat \imath}}
  = & -\int_0^T d X^0 \int \frac{d^d \pi}{(2 \pi \hbar)^d} \sum_n \notag \\
  & \times [\Omega_{n {\vec \pi} X^0}^{\hat \imath} (\epsilon_{n {\vec \pi} X^0} - \mu) - m_{n {\vec \pi} X^0}^{\hat \imath}/2]
  f_{n {\vec \pi} X^0}, \label{eq:dpqc1a} \\
  \Delta P_{\rm Qc}^{(1b) {\hat \imath}}
  = & \frac{1}{2} \int_0^T d X^0 \int \frac{d^d \pi}{(2 \pi \hbar)^d} \sum_n \notag \\
  & \times m_{n {\vec \pi} X^0}^{\hat \imath} f^{\prime}_{n {\vec \pi} X^0} (\epsilon_{n {\vec \pi} X^0} - \mu), \label{eq:dpqc1b} \\
  \Delta P_{\rm Qc}^{(1a) {\hat \imath}}
  = & -\frac{1}{2} \int_0^T d X^0 \int \frac{d^d \pi}{(2 \pi \hbar)^d} \sum_n \notag \\
  & \times m_{n {\vec \pi} X^0}^{\hat \imath} f^{\prime}_{n {\vec \pi} X^0} (\epsilon_{n {\vec \pi} X^0} - \mu). \label{eq:dpqc1c}
\end{align} \label{eq:dpqc1} \end{subequations}
Again, Eqs.~\eqref{eq:dpqc1b} and \eqref{eq:dpqc1c} are canceled exactly.

The remaining term Eq.~\eqref{eq:dpqc1a} is generally nonzero even at zero temperature, and hence is not physically correct.
Indeed, there is a correction term arising from the heat generation.
Such heat current is defined by
\begin{equation}
  J_{\rm Qnc}^{\hat \imath}
  \equiv \frac{\partial \Omega}{\partial h_{\hat \imath}^{\phantom{\hat \imath} 0}}, \label{eq:jqnc}
\end{equation}
where $\Omega \equiv E - T S - \mu N$ is the free energy.
However, since the free energy is difficult to calculate directly, the total energy $K \equiv E - \mu N$ is calculated below,
\begin{equation}
  K(X)
  = \pm \frac{ i \hbar}{2} \int \frac{d^D \pi}{(2 \pi \hbar)^D} \tr [(-\pi_{\hat 0}) \star {\hb G} + {\hb G} \star (-\pi_{\hat 0})]^<. \label{eq:k}
\end{equation}
Owing to symmetrization, the star product is reduced to the ordinary product, and the auxiliary heat current is obtained as
\begin{align}
  {\tilde J}_{\rm Qnc}^{\hat \imath}
  \equiv & \frac{\partial K}{\partial h_{\hat \imath}^{\phantom{\hat \imath} 0}} \notag \\
  = & \pm \frac{i \hbar^2}{2} \int \frac{d^D \pi}{(2 \pi \hbar)^D} f(-\pi_{\hat 0}) (-\pi_{\hat 0})
  \tr G_{1, h_{\hat \imath}^{\phantom{\hat \imath} 0}}^{< (0)}, \label{eq:auxjqnc}
\end{align}
where
\begin{subequations} \begin{align}
  G_{1, h_{\hat \imath}^{\phantom{\hat \imath} 0}}^{\rm R}
  = & G_0^{\rm R} \Sigma_{1, h_{\hat \imath}^{\phantom{\hat \imath} 0}}^{\rm R} G_0^{\rm R} \notag \\
  & + i (G_0^{\rm R} \partial_{X^0} G_0^{{\rm R} -1} G_0^{\rm R} \partial_{\pi_{\hat \imath}} G_0^{{\rm R} -1} G_0^{\rm R}
  - (X^0 \leftrightarrow \pi_{\hat \imath})), \label{eq:g1hr} \\
  G_{1, h_{\hat \imath}^{\phantom{\hat \imath} 0}}^{< (0)}
  = & \pm (G_{1, h_{\hat \imath}^{\phantom{\hat \imath} 0}}^{\rm R} - G_{1, h_{\hat \imath}^{\phantom{\hat \imath} 0}}^{\rm A}), \label{eq:g1h0} \\
  \Sigma_{1, h_{\hat \imath}^{\phantom{\hat \imath} 0}}^{< (0)}
  = & \pm (\Sigma_{1, h_{\hat \imath}^{\phantom{\hat \imath} 0}}^{\rm R} - \Sigma_{1, h_{\hat \imath}^{\phantom{\hat \imath} 0}}^{\rm A}), \label{eq:s1h0}
\end{align} \label{eq:perturbxh} \end{subequations}
and $G_{1, h_{\hat \imath}^{\phantom{\hat \imath} 0}}^{< (1)} = \Sigma_{1, h_{\hat \imath}^{\phantom{\hat \imath} 0}}^{< (1)} = 0$.
To translate the auxiliary heat current to the proper one, it is necessary to solve the differential equation,
\begin{equation}
  \frac{\partial (\beta_0 J_{\rm Qnc}^{\hat \imath})}{\partial \beta_0}
  = {\tilde J}_{\rm Qnc}^{\hat \imath}. \label{eq:transjqnc}
\end{equation}
In the clean and non-interacting limit, we get
\begin{align}
  {\tilde J}_{\rm Qnc}^{\hat \imath}
  = & \frac{\hbar^2}{2} \int \frac{d^D \pi}{(2 \pi \hbar)^D} f(-\pi_{\hat 0}) (-\pi_{\hat 0}) \notag \\
  & \times \tr [g_0^{\rm R} v^{\hat \imath} g_0^{\rm R} {\dc H} g_0^{\rm R} - ({\rm R} \to {\rm A})] - (v^{\hat \imath} \leftrightarrow {\dc H}) \notag \\
  = & \frac{1}{2} \int \frac{d^D \pi}{(2 \pi \hbar)^D} \sum_n
  [2 \Omega_{n {\vec \pi} X^0}^{\hat \imath} f_{n {\vec \pi} X^0} (\epsilon_{n {\vec \pi} X^0} - \mu) \notag \\
  & - m_{n {\vec \pi} X^0}^{\hat \imath} (f_{n {\vec \pi} X^0} + f^{\prime}_{n {\vec \pi} X^0} (\epsilon_{n {\vec \pi} X^0} - \mu))], \label{eq:auxjqnc1}
\end{align}
By solving Eq.~\eqref{eq:transjqnc} and performing the temporal intergral, the heat-generation contribution to the change in the HP is given by
\begin{align}
   \Delta P_{\rm Qnc}^{\hat \imath}
   \equiv & \int_0^T d X^0 J_{\rm Qnc}^{\hat \imath} \notag \\
   = & -\frac{1}{2} \int_0^T d X^0 \int \frac{d^d \pi}{(2 \pi \hbar)^d} \sum_n \notag \\
   & \times \left[m_{n {\vec \pi} X^0}^{\hat \imath} f_{n {\vec \pi} X^0}
  + 2 \Omega_{n {\vec \pi} X^0}^{\hat \imath} \int_{\epsilon_{n {\vec \pi} X^0} - \mu}^{\infty} d z f(z)\right]. \label{eq:dpqnc}
\end{align}

After all, the change in the HP consists of Eqs.~\eqref{eq:dpqc1} and \eqref{eq:dpqnc} as
\begin{align}
  \Delta P_{\rm Q}^{\hat \imath}
  = & -\int_0^T d X^0 \int \frac{d^d \pi}{(2 \pi \hbar)^d} \sum_n \notag \\
  & \times \Omega_{n {\vec \pi} X^0}^{\hat \imath}
  \left[f_{n {\vec \pi} X^0} (\epsilon_{n {\vec \pi} X^0} - \mu) + \int_{\epsilon_{n {\vec \pi} X^0} - \mu}^{\infty} d z f(z)\right]. \label{eq:dpq}
\end{align}
In the case of a fermion, the Sommerfeld expansion holds at low temperature,
\begin{equation}
  f^{\prime}(z)
  = -\delta(z) - \frac{\pi^2 T_0^2}{6} \delta^{\prime \prime}(z) + \dots, \label{eq:sommerfeld}
\end{equation}
and the change in the HP is approximated by
\begin{align}
  \Delta P_{\rm Q}^{\hat \imath}
  = & \int_0^T d X^0 \int \frac{d^d \pi}{(2 \pi \hbar)^d} \sum_n
  \Omega_{n {\vec \pi} X^0}^{\hat \imath} \int_{\epsilon_{n {\vec \pi} X^0} - \mu}^{\infty} d z f^{\prime}(z) z \notag \\
  = & -\frac{\pi^2 T_0^2}{3} \int_0^T d X^0 \int \frac{d^d \pi}{(2 \pi \hbar)^d} \sum_n 
  \Omega_{n {\vec \pi} X^0}^{\hat \imath} \delta(\mu - \epsilon_{n {\vec \pi} X^0}) \notag \\
  = & \frac{\pi^2 T_0^2}{3 q} \frac{\partial \Delta P^{\hat \imath}(T_0 = 0)}{\partial \mu}. \label{eq:mott}
\end{align}
This is analogous to the Mott rule for the thermoelectric conductivity.
In fact, if we identify the adiabatic parameter $X^0$ to the momentum in the extra dimension,
the heat-transfer contribution Eq.~\eqref{eq:dpqc1} is equivalent to
the Kubo-formula contribution to the thermoelectric Hall conductivity multiplied by the temperature,
and the heat-generation contribution Eq.~\eqref{eq:dpqnc} is equivalent to the orbital magnetization~\cite{PhysRevB.86.214415}.
As a result, the HP always vanishes not only at zero temperature as expected, but also near zero temperature for inversion-broken insulators whose Fermi energy lies in the gap.
The analogy between charge and heat does not always hold, at least in the polarization.

%--- Discussion
As seen above, the HP does not appear spontaneously even in inversion-broken insulators,
but can be induced by external fields.
Especially, when we apply a torsional magnetic field $T^{\hat 0}_{\phantom{\hat 0} {\hat k} {\hat l}}$,
which is coupled to the energy $-\pi_{\hat 0}$ in the Poisson bracket Eq.~\eqref{eq:poisson},
we expect the Wiedemann-Franz law instead of the Mott rule.
As a result, in $(3+1)$-D TIs/TSCs, the HP can be induced as
\begin{equation}
  \frac{1}{2} \epsilon_{{\hat \jmath} {\hat k} {\hat l}} \frac{\partial P_{\rm Q}^{\hat \imath}}{\partial (-T^{\hat 0}_{\phantom{\hat 0} {\hat k} {\hat l}})}
  = \frac{\pi^2 T_0^2}{3 q^2} \frac{1}{2} \epsilon_{{\hat \jmath} {\hat k} {\hat l}} \frac{\partial P^{\hat \imath}}{\partial F_{{\hat k} {\hat l}}}
  = \frac{\theta T_0^2}{12 \hbar} \delta^{\hat \imath}_{\phantom{\hat \imath} {\hat \jmath}}. \label{eq:cross}
\end{equation}
Here the orbital magneto-electric susceptibility
$\partial P^{\hat \imath}/\partial B^{\hat \jmath} = \theta q^2 \delta^{\hat \imath}_{\phantom{\hat \imath} {\hat \jmath}}/4 \pi^2 \hbar$
results from the axion electrodynamics in $(3+1)$-D TIs with $\theta = \pi$
~\cite{PhysRevB.78.195424,PhysRevLett.102.146805,PhysRevB.81.205104,1367-2630-12-5-053032,PhysRevB.84.205137},
and an extra factor $1/2$ is multiplied in $(3+1)$-D TSCs owing to their Majorana nature~\cite{PhysRevLett.108.026802}.
Such cross correlation can be encoded in the effective action,
\begin{equation}
  S_{\rm eff}
  = -\frac{\theta T_0^2}{96 \hbar} \int d^4 X
  \epsilon^{\mu \nu \rho \sigma} T^{\hat 0}_{\phantom{\hat 0} \mu \nu} T^{\hat 0}_{\phantom{\hat 0} \rho \sigma}, \label{eq:action}
\end{equation}
which is the temporal part of the Nieh-Yan action
$(T^a \wedge T_a - R^{ab} \wedge e_a \wedge e_b)/4 \pi^2 l^2 = d (e^a \wedge T_a)/4 \pi^2 l^2$
with the dimensional parameter $l^{-1} \propto T_0$~\cite{nieh:373}.
It is noted that the spatial part, which may have a different dimensional parameter,
describes the topological viscoelastic responses~\cite{PhysRevLett.107.075502,Hidaka01012013,PhysRevD.88.025040}.
Since we do not assume the Lorentz symmetry, the temporal and spatial dimensional parameters do not necessarily coincide.
The important point to obtain Eq.~\eqref{eq:action} is that the heat current has a corresponding gauge potential $h^{\hat 0}_{\phantom{\hat 0} i}$,
and hence the HP is coupled to a torsional electric field $T^{\hat 0}_{\phantom{\hat 0} i0}$.
Here a torsional electric field $T^{\hat 0}_{\phantom{\hat 0} i0}$ is induced by a gravitational potential $h^{\hat 0}_{\phantom{\hat 0} 0}$,
and is the most convenient representation of a gravitational field~\cite{1310.8043}.
If we assume the torsion-free condition, a spin connection is another representation.
However, it is a natural description that displacement responses are induced by gauge fields but not by gauge potentials.
As Luttinger proposed, a gravitational field is the mechanical force equivalent to a temperature gradient~\cite{PhysRev.135.A1505}.
On the other hand, a torsional magnetic field $T^{\hat 0}_{\phantom{\hat 0} ij}$ coupled to the heat magnetization
is different from an angular velocity of rotation~\cite{1310.8043}.
Its physical realization, which is also related to the direct measurement of the heat magnetization, is one important problem.
Another important problem is the microscopic derivation of Eq.~\eqref{eq:cross},
which will be accomplished by the second-order perturbation theory with respect to the temporal gradient and a torsional magnetic field.

Finally, let us comment on some advantages of our Keldysh formalism.
Although the HP, as well as the CP, is well-defined only near zero temperature,
it is properly derived by calculating the heat current at finite temperature and then using the Sommerfeld expansion.
Furthermore, our formalism can be applied to disordered or interacting systems within the perturbation theory.
The Matsubara formalism, which is equivalent to ours, was already used to calculate the CP in combination with the dynamical mean-field theory~\cite{PhysRevB.88.155121}.
On the other hand, the Berry-phase formalisms for the CP~\cite{PhysRevB.49.14202,PhysRevLett.80.1800} are practically difficult
since they need the ground-state wave-function.

%--- Summary
To summarize, we have shown systematic derivation of the CP/HP by using the Keldysh formalism.
We have found that the HP consists of the heat-transfer and heat-generation contributions, and vanishes at low temperature in inversion-broken insulators.
This is the first counterexample of the charge-heat analogy we naively expected.
We have also proposed that the heat cross correlation possible in $(3+1)$-D TIs/TSCs is described by the temporal part of the Nieh-Yan action.

%--- Acknowledgement
\begin{acknowledgements}
  This work was supported by Grant-in-Aid for Japan Society for the Promotion of Science Fellows No.~$24$-$600$.
\end{acknowledgements}
%--- References

\end{document}